\documentclass[preprint]{iucr}              
\usepackage {amsmath}									
\usepackage {amsfonts}									

\begin{document}                  



\title{Generation of basis vectors and the importance of unitary irreducible representations}


\author[a]{Z. L.}{Davies}
\cauthor[a]{A. S.}{Wills}{a.s.wills@ucl.ac.uk}{}

\aff[a]{University College London, Department of Chemistry, 20 Gordon Street, London, WC1H 0AJ \country{England}}









\maketitle                        

\begin{synopsis}
Restraints and techniques in generating symmetry adapted functions for crystallography.
\end{synopsis}

\begin{abstract}
Increasing attention is being focused on the used of symmetry-adapted functions to describe magnetic structures, structural distortions, and incommensurate crystallography. Though the calculation of such functions is well developed, significant difficulties can arise, such as the generation of too many or to few functions to minimally span all of a systems degrees of freedom. We present an elegant solution to these difficulties using the concept of basis sets, and discuss previous work in this area using this concept. Further, we highlight the significance of unitary irreducible representations in this method, and provide the first validation that the irreducible representations of the crystallographic space groups presented by Kovalev are unitary.
\end{abstract}


\section{Introduction}

The use of symmetry adapted functions is well established in many fields, such as electronic structure calculations and vibrational mode analysis, increasingly its value to the crystallographer is being recognized too. The use of such functions in the refinement of magnetic structures is already supported in SARA\emph{h}-GSAS (Wills, 2000) and Fullprof (Rodriguez-Carvajal, 1993), and application of this technique to atomic structure is an area of current interest and development: Campbell \emph{et. al.} (2007), Wills (2008). Analysis using these functions has been applied to several neutron scattering techniques: spherical neutron polarimetry of Er$_{2}$Ti$_{2}$O$_{7}$ (Poole \emph{et al.}, 2007); single crystal diffraction of CuB$_{2}$O$_{4}$ (Boehm \emph{et al.}, 2003); and powder diffraction of dicyanamide-based molecular magnets (Lappas \emph{et. al}, 2003). Recent interest has focused upon applications of this technique to displacive phase transitions: Wills (2005), Campbell \emph{et al.} (2007), Wills and Davies (2008), and Perez-Mato \emph{et al.} (2008).

Symmetry adapted functions are derived using the techniques of representation theory, and are intimately related to the eigenfunctions of the Hamiltonian (Bradley, Cracknell 1972, Izyumov 1991). In crystallography these functions are usually referred to as basis vectors (BVs) and they define the order parameters of some property; the coefficients of the BVs define the state of a system. The eigenfunctions of a spin Hamiltonian have been used in an analogous manner to basis vectors for magnetic systems (Bertaut, 1962 \& 1981).

Constructing BVs is a simple, though laborious, task (Izyumov, 1990). A representation of some system property under its symmetry operations is constructed as the direct product of two simpler representations $\Gamma = \Gamma_{V} \otimes \Gamma_{Perm}$. In this notation, $\Gamma_{Perm}$ represents the permutation of atoms under the symmetry operations of the system; $\Gamma_{V}$ represents how some property of interest transforms under the same operations. Fortunately, the full matrix representation is not required and the character of the matrices is sufficient. Once these have been determined, $\Gamma$ is reduced to a linear combination of irreducible representations (IRs): $\Gamma = \sum_{i}C^{i}\Gamma^{i}$. The final step is the generation of BVs using the method of projection operators. 

A common problem when performing these calculations is the projection of too few or too many basis vectors. Under-generation has been thoroughly explored by Stokes \emph{et al.} (1991), however, the problem of over-generation has not been previously resolved. In this work we discuss projection operators using the concept of ``basis sets'' and apply use concept to resolve the over- and under-generation; we illustrate the general procedure with an example. Further, in Section 7 we provide the first confirmation (that the authors are aware of) that the IRs listed in the tables of Kovalev (1993) are unitary and the highlight importance of this property. We begin by defining what constitutes an appropriate set of solutions, and how they are derived.

\section{Properties of Basis Sets}

Any property of a crystal, such as atomic displacements or magnetic moment, is described by a field. The axis system of such a field is arbitrary, and the system and is divisible into subspaces, each with the symmetry of one IR in the linear expansion $\Gamma = \sum_{i}C^{i}\Gamma^{i}$. Each subspace is spanned by sets of BVs with appropriate symmetry: \emph{basis sets}. The symmetric basis of the whole system is given by the set of all basis sets. The desired basis for the subspace which transforms as $\Gamma^{\nu}$ is a set of  vectors $\{ \psi^{\nu}_{1}, \psi^{\nu}_{2}, \ldots, \psi^{\nu}_{d^{\nu}}\}$ for which:
\begin {equation} \label {Eq:BVProperty}
g \psi^{\nu}_{l} = \sum^{d^{\nu}}_{m=1} \mathfrak{d}^{\nu}_{ml}(g) \psi^{\nu}_{m}, \qquad \forall g \in \mathbb{G}
\end {equation}
Here $\mathbb{G}$ is the symmetry group of the system, $\mathfrak{d}^{\nu}(g)$ is the matrix representing the operation $g$ in the IR $\Gamma^{\nu}$, $d^{\nu}$ is the order of $\Gamma^{\nu}$. A set of BVs obeying Eq. \ref{Eq:BVProperty} will be referred to as a \emph{basis set}; note that basis sets \emph{always} occur in sets of order $d^{\nu}$. We will use the properties of basis sets to explore over- and under-generation in sections 4 and 5.

\section{Presentation of the Operators}

Here we present the key equations in the method of projection operators in two forms (Hammermesh, 1964), and discuss the importance of unitary IRs when using the standard forms. The reduction formula is defined as:
\begin {equation} \label{EqReduction}
\frac{1}{|\mathbb{G}|} \sum_{g_{s} \in \mathbb{G}} \chi^{\nu}(g_{s})\chi^{\mu}(g_{s}^{-1})=  \delta_{\nu, \mu}
\end{equation}
Where $\chi^{n}(g_{s})$ is the character of the matrix representing $g_{s}$ in the representation $\Gamma^{n}$. The action of the reduction operator is to determine the coefficients, $C^{\mu}$, when applied to some representation $\Gamma^{\mu}= \sum_{\mu}C^{\mu}\Gamma^{\mu}$. When a matrix is the conjugate transpose of its inverse, then it is said to be \emph{unitary}; an IR composed of unitary matrices is, itself, unitary.  Under the assumption of unitary IRs Eq. \ref{EqReduction} can be rewritten in its usual form:
\begin{eqnarray}
\mathfrak{d}_{ml}(g_{s})&=&\mathfrak{d}_{lm}^{*}(g_{s}^{-1})\\
\frac{1}{|\mathbb{G}|} \sum_{g_{s} \in \mathbb{G}} \chi^{\nu}(g_{s})\chi^{\mu *}(g_{s})&=&  \delta_{\mu, \nu}
\end {eqnarray}

Similarly there are two forms of the projection operator\footnote{Strictly, the operators presented in Eqs. 5 and 6 are not projection operators; this name is reserved for their idempotent derivatives for which $l=m$. However, the phrase ``method of projection operators" is synonymous with technique and therefore we shall refer them by this name from herein.}. The first of these is applicable to any IR, while the second line is the projection operator derived under the assumption of unitary IRs:
\begin{eqnarray}\label{Eq:Pro}
W^{\mu}_{ml}\psi^{\nu}_{i} &=& \frac{d^{\mu}}{|\mathbb{G}|}\sum_{g_{s} } \mathfrak{d}^{\mu}_{ml}(g_{s}^{-1}) \mathfrak{T}(g_{s})\psi^{\nu}_{i}
 \\
W^{\mu}_{lm} &=& \frac{d^{\mu}}{|\mathbb{G}|}\sum_{g_{s}} \mathfrak{d}^{\mu*}_{lm}(g_{s}) \mathfrak{T}(g_{s}) \\
W_{lm}^{\mu}\psi_{m}^{\mu} &=& \psi_{l}^{\mu}
\end {eqnarray}

The projection operator $W^{\mu}_{lm}$ generates a set of vectors of order $d^{\mu}$ from a single vector by permuting the BVs amongst themselves; specifically a group of BVs the symmetry of $\Gamma^{\mu}$. Thus, the projection operator will ``project out'' a complete basis set from a single BV. However, in general we will not know any of the BVs, and we must consider the action of the projection operator upon a trial vector, $\phi = \sum_{\nu}\sum^{d^{\nu}}_{i}C^{\nu}_{i}\psi^{\nu}_{i}$:
\begin {eqnarray}
W_{lm}^{\mu}\phi &=& \sum_{\nu}\sum^{d^{\nu}}_{i} C^{\nu}_{i} W_{ml}^{\mu}\psi^{\nu}_{i} \nonumber\\
&=& \sum_{\nu}\sum^{d^{\nu}}_{i} C^{\nu}_{i}\psi_{l}^{\mu}\delta_{m,i}\delta_{\mu,\nu} \nonumber\\
&=& C^{\mu}_{m}\psi^{\mu}_{l}
\end {eqnarray}
The action of the projection operator is to take the component of $\phi$ along $\psi^{\mu}_{m}$ and transforms it into $\psi^{\mu}_{l}$; all the other components are transformed to zero. The projection operator is most commonly encountered in the form of Eq. 6, which restricts the IRs to being unitary; we discuss this restriction in section 7.

\section{Over-generation}

Over-generation is the projection of more BVs than required to minimally span a system's degrees of freedom. Correspondingly, some of the derived BVs are linearly related. The problem is how best to reduce our set of solutions to one that is minimal while preserving the symmetry properties required for our basis.

Often equivalent solutions will occur in pairs, related by a complex coefficient: $\psi_{1} =C\psi_{2}$. In this case it is simple for an algorithm to determine the equivalence relationships. However, when three or more BVs are linearly related then, because our BVs are not required to be orthogonal, there is no simple way to determine linear relationships between them. In particular, there is no simple algorithm to determine equivalence relations and thus calculate which BVs are equivalent. We face a further problem: only entire basis sets can be eliminated, otherwise our solution is not a set of basis sets and cannot have the correct symmetry properties.

As the projection operator must derive a set of order $d^{\nu}$ or $0$ for each trial, it is apparent that when over-generation occurs too many trial functions have been used. If, through judicious choice of trial vectors, we can reduce all BV equivalencies to the form $\psi_{1} =C\psi_{2}$, then it becomes possible to determine which trial vectors give equivalent answers and eliminate all but one of them. By eliminating trial-vectors, and not basis vectors, the symmetry properties of the BVs are preserved.

\subsection {Symmetry Adapted Trial vectors}

Our goal is to determine a method of constructing an appropriate set of trial functions. If two BVs are linearly related then the system property at each point must have the same linear relationship. Thus, by controlling how the property at a single point is generated we can control the generation of the entire BV. Under the methord of projection operators, the property at a single point is generated by the sum action of all the operators which generate that point from an initial position $A_{0}$; if we consider $A_{0}$ itself then these operators are the co-called ``stabilizers'' of $A_{0}$, denoted $\mathbb{H}_{0}$.
\begin {equation}
\begin {split}
\mathbb{H}_{0} &\subset \mathbb{G}_{k}\\
\frac{d^{\mu}}{|\mathbb{G}|}\sum_{h_{s} \in \mathbb{H}_{0} } \mathfrak{d}^{\mu}_{ml}(h_{s}^{-1}) \mathfrak{T}(g_{s})\phi^{\nu}_{i}
&=
\left( \begin {array}{c} a_{0}\\b_{0}\\c_{0} \end{array} \right) A_{0}
\end {split}
\end {equation} 
The property at atom $A_{0}$ is defined by the vector ($a_{0}$, $b_{0}$, $c_{0}$), defined by the crystallographic axes. Izyumov (1990) has developed a formalism for the reduction and projection operators using stabilizers. 

The stabilizers of $A_{0}$ form a group and thus subdivide the $R^{3}$ space into invariant subspaces, by selecting our trial vectors to lie within lines and planes of invariance we naturally simplify the relations between our projected BVs. In many cases it will be obvious where the lines and planes of invariance lie for a stabilizer group, when it is not we can construct them using the method of projection operators twice. First, the trial functions are themselves projected using $\mathbb{H}_{0}$ and the trial vectors $\phi_{1}$=$(1,0,0)$, $\phi_{2}$=$(0,1,0)$, $\phi_{3}$=$(0,0,1)$ at the position $A_{0}$. These symmetry-adapted trial functions can then be used to perform the projection of the system's BVs.

This technique is particularly appropriate when the lowering of a systems symmetry divides related positions into a number of orbits. Consider some position $A_{i} = g_{i}A_{0}$, where $g_{i}$ consists of a rotation-reflection $h_{i}$ and a translation $\tau_{i}$. If $\mathbb{H}_{i}$ is the group of operations ``stabilizing'' $A_{i}$ then $\mathbb{H}_{i} = g_{i}\mathbb{H}_{0}g_{i}^{-1}$. Thus, if two orbits are related by the operation $g_{i}$, appropriate set of trial functions $\psi_{trial}$ are related by $h_{i}\psi_{trial}$. 
\begin {equation}
\begin {split}
\mathbb{H}_{i} &= g_{i}\mathbb{H}_{0}g_{i}^{-1}\\
\psi_{trial, orbit_{i}} &= h_{i}\psi_{trial, orbit_{0}}
\end{split}
\end {equation}

In the next section we will work through an example where the standard trial functions produce an excess of solutions, and determine a more appropriate trial set. The example splits into two orbits under the distorted ordering, and we show that the trial-functions for the both orbits have the relationship given in Eq. 10.

\subsection*{Worked Example}

Consider the space group $I4_{1}32$ (214), and the $k$-vector $k=(\frac{1}{2}, \frac{1}{2}, \frac{1}{2})$, with an atom at the position $(0,0,0)$. Under the operations of the space group of the $k$-vector, $G_{k}$, there are three equivalent positions at $(\frac{1}{2}, \frac{1}{2}, 0)$, $(0, \frac{1}{2}, \frac{1}{2})$, and $(\frac{1}{2}, 0, \frac{1}{2})$. Using SARA\emph{h}, the decomposition of possible atomic displacements is given as:
\begin {equation}
\Gamma_{Mag} = 2\Gamma^{1} + 2\Gamma^{2} + 2\Gamma^{3}
\end {equation}
Every IR is of order $2$, and therefore we expect $2\times2=4$ BVs to be projected from each IR. 

The basis vectors generated for $\Gamma_{1}$ using the standard trial vectors $\phi_{1}$=$(1,0,0)$, $\phi_{2}$=$(0,1,0)$, $\phi_{3}$=$(0,0,1)$ are listed in Table 1, using the notation:
\begin {equation}
\psi^{n}_{ij}(x,y,z) = \left( \begin {array}{c} a_{0}\\b_{0}\\c_{0} \end{array} \right) A_{0} + ...
\end {equation}
The BV $\psi^{n}_{ij}(x,y,z)$ has been projected from the IR $\Gamma^{n}$, using the $ij^{th}$ matrix element of each matrix, and the trial vector $(x, y, z)$ at the position $A_{0}=(0,0,0)$. It consists of a series of vectors $(a_{n}, b_{n}, c_{n})$, defined with respect to the crystallographic axes, at the positions $A_{n}$. Projection using the standard trial functions generates six apparently distinct BVs, rather than the four required by the reduction formula, clearly one of the trial functions is superfluous. It can be shown that $\psi^{1}_{11}(0,0,1) = e^{\frac{2}{3}\pi.i}\psi^{1}_{11}(1,0,0) + e^{-\frac{2}{3}\pi.i}\psi^{1}_{11}(0,1,0)$, however, this solution neither apparent on inspection, nor simple to determine.

Following the strategy of section 4.2, we seek to determine a set of symmetry-adapted trial functions to simplify the BV relationships. The stabilizer group of the position $A_{0}$ is the group of $C_{3}$ rotations about the $(1,1,1)$. The invariant subspaces of this group are the line $(1,1,1)$ and the perpendicular plane $[1,1,1]$. Thus we select one trial vector to lie along $(1,1,1)$ and the other two to lie in $[1,1,1]$ chosen to form a right-hand set: $\phi_{1}$=$(1,1,1)$, $\phi_{2}$=$(1,-1,0)$, $\phi_{3}$=$(1,1,2)$. As presented in table 2, these trial vectors have been renormalized to have modulus 1. Also listed are the BVs generated from the symmetry-selected set of trial functions. By inspection, $\frac{1}{\sqrt{2}}\psi^{1}_{11}(1,-1,0) = -i.\frac{1}{\sqrt{6}}\psi^{1}_{11}(1,1,-2)$, and we can eliminate either $\phi_{2}=(1,-1,0)$ or $\phi_{3}=(1,1,-2)$ from our projection.  Thus by suitable selection of trial functions we have produced a set of BVs in which excess solutions are readily discernible.

Further, our example is split into two orbits, the second orbit being related to the previously considered set of atomic positions by the operation:
\begin {equation}
\begin{split}
g_{5} &= \left(
\begin{array}{cccc} 
0 & 1 & 0 & 0.25\\
1 & 0 & 0 & 0.75\\
0 & 0 & -1 & 0.75\\
0 & 0 & 0 & 1 
\end{array}
\right)\\
h_{5} &= \left(
\begin{array}{ccc} 
0 & 1 & 0\\
1 & 0 & 0\\
0 & 0 & -1 
\end{array}
\right)
\end{split}
\end {equation}

Using Eq. 10, the following trial functions are generated for the second orbit: $\phi_{1}$=$(1,1,-1)$, $\phi_{2}$=$(-1,1,0)$, $\phi_{3}$=$(1,1,2)$ should generate BVs with the desired linear relationships. The BVs for the \emph{second} orbit, using this set of trial functions, are presented in Table 3. Inspection reveals that $\frac{1}{\sqrt{2}}\psi^{1}_{11}(-1,1,0) = i.\frac{1}{\sqrt{6}}\psi^{1}_{11}(1,1,2)$, and again we are free to eliminate either $\phi_{2}$ ot $\phi_{3}$. We note that this method is also applicable to orbits joined under co-representations. 

\section{Under-generation}

Undergeneration is the apparent inability to fully span a systems decomposition using the BVs generated by the method of projection operators. For the projection operator $W^{\mu}_{ij}$ changing $\mu$ generates a basis set with a different symmetry, hence the only free variable with which to resolve under-generation is $ij$. This problem has been thoroughly explored by Stokes \emph{et. al} (1991), who define when varying the column index $j$ will generate inequivalent BVs. It is useful to discuss this problem using basis sets to demonstrates the power of this concept in understanding the method of projection operators.

Basis vectors occur in basis sets which transform under two relations:
\begin {equation}
\begin {split}
g\psi_{i}^{\mu} &= \sum_{j}^{d^{\mu}}\mathfrak{d}^{\mu}_{ji}(g) \psi^{\nu}_{j}\\
W^{\mu}_{ij}\psi_{j}^{\mu} &= \psi^{\mu}_{i}
\end {split}
\end {equation}
It is apparent from consideration of these two equations that the enumeration of BVs is not arbitrary; it defines how BVs inter-relate \emph{within the basis set to which they belong}. Further, the number of basis sets of a symmetry $\Gamma^{\mu}$ is exactly $C^{\mu}$, and within each set the BV's will be labelled $1, 2, ..., d^{\nu}$. Thus, while the numbering is not arbitrary it is not unique either.  The action of $W_{i1}^{\mu}$ on a general vector $\psi$ is to project the component along $\psi_{1}^{\mu}$ into $\psi^{\mu}_{i}$. Similarly, the action of $W_{i2}^{\mu}$ is to project the component along $\psi_{2}^{\mu}$ into $\psi^{\mu}_{i}$. However, there is no restriction that $\psi_{1}$ and $\psi_{2}$ are from the same basis set. 

Thus, we conclude that varying the row-index $i$ generates another member of the same basis-set, while varying the column-index $j$ generates a BV from a different basis set (which may be equivalent).

\section{Unitary Check}

In section 3 it was emphasised that the projection and reduction operators are normally encountered in a form which restricts the IRs to being unitary. Their use with non-unitary IRs would generate BVs lacking the correct symmetry properties. The absence of symmetry relations between the BVs could be realized as under- or over- generation, therefore it is key to discuss possible sources of IRs for these calculations.

IRs derived from Zak's method (Zak, 1960) are, by derivation, unitary; the IRs output by the computer codes KAREP (Aroyo, 1992) and REPRES (Aroyo, 2003) are examples of such. However, when using IRs from collated tables the unitary properties must be confirmed explicitly. As part of this work we have verified the tables of Kovalev (1991) a standard reference for the IRs of crystallographic space groups; this verification has been absent until now (Davies 2008). 

The method of verification was brute-force calculation. Our algorithm determined for each symmetry element $g_{1}$, some symmetry element $g_{2}$ for which $g_{1}g_{2}$ is an identity-translation, represented by a complex number $C$. If the IR is a unitary homomorphism then $C^{*}$ will transform $\mathfrak{d}(g_{2})$ into the conjugate-transpose of $\mathfrak{d}(g_{1})$ :
\begin {equation}
\begin {split}
&\mathfrak{T}(g_{1}) \times \mathfrak{T}(g_{2}) = 
\begin{array}{|cccc|}
1&0&0&T_{x}\\
0&1&0&T_{y}\\
0&0&1&T_{z}\\
0&0&0&1\\
\end{array} \\
&\mathfrak{d}^{\dagger}(g_{1}) = \mathfrak{d}(g_{2}) \times exp \left(-2\pi\times(k_{x}, k_{y},k_{y}) \cdot \left(\begin {array} {c} T_{x}\\T_{y}\\T_{z}\\ \end{array} \right) \right)  
\end{split}
\end {equation}
The matricies $\mathfrak{T}(g_{i})$ are the normal $4\times4$ matrix representation of the affine operation $g_{i}$. The vector $(k_{x}, k_{y},k_{y})$, termed the $k$-vector, defines the translational periodicity of a modulated structure, and the representation of a translation is a funtion of its dot-product with the $k$-vector.

This work showed that the IRs presented in Kovalev's tables are indeed unitary and validate the projection and reduction techniques used in the computer codes based upon them such as SARA\emph{h} (Wills, 2000), MODY (Sikora, 2004) and Isotropy (Stokes, 2007). 

\section{Conclusions}

A common problem in the application of projection techniques to physical problems in crystallography is the over- and -under generation of basis vectors. Understanding the method in terms of basis sets allows a solution to over-generation to be constructed through use, and subsequent elimination, of symmetry adapted trial functions. This technique ensures that the solution is a set of basis sets and has all the required symmetry properties. 

Further we have verified the unitary nature of the IRs presented by Kovalev, and shown that any observed difficulties in the projection of a minimal spanning set are not a consequence of failing to take account of restrictions placed upon the IRs by the projection operator.



\ack{Acknowledgements}

The authors would like to aknowledge the ESRC for funding this work.



\begin{center}
\begin{table*}[ht]
\caption{The BVs projected for our example system using the crystallographic axes as trial functions.}
{\small
\hfill{}
$\begin{array}{l|cccc}      
BV & A_{0} = (0,0,0) & A_{1} = (\frac{1}{2}, \frac{1}{2}, 0) &A_{2} = (0, \frac{1}{2}, \frac{1}{2})& A_{3} = (\frac{1}{2}, 0, \frac{1}{2}) \\
\hline
&&&&\\
\psi^{1}_{11}(1,0,0) & 
\left(\begin{array}{c}1\\-0.183 -	0.683i\\ -0.183 + 0.683i \end{array}\right) &
\left (\begin{array}{c}0 + i\\-0.683 +	0.183i\\ 0.683 + 0.183i\\ \end{array}\right) &
\left (\begin{array}{c}0\\0.683 - 0.183i\\ -0.183 + 0.683i \end{array}\right)&
\left (\begin{array}{c}0\\0.183 +	0.683i\\ -0.683 - 0.183i \end{array}\right) \\
&&&&\\
\psi^{1}_{11}(0,1,0) &
\left (\begin{array}{c}-0.183 +	0.683i\\1\\ -0.183 - 0.683i \end{array}\right)&
\left (\begin{array}{c}-0.683 - 0.183i\\0 - i\\ -0.683 + 0.183i \end{array}\right)&
\left (\begin{array}{c}-0.183 +	0.683i\\0\\ -0.683 + 0.183i \end{array}\right)&
\left (\begin{array}{c}0.683 +	0.183i\\0\\ -0.183 - 0.683i\\ \end{array}\right)\\
&&&&\\
\psi^{1}_{11}(0,0,1) & 
\left (\begin{array}{c}-0.183	- 0.683i\\-0.183 +	0.683i\\ 1 \end{array}\right)&
\left (\begin{array}{c}0.683 - 0.183i\\0.683 +	0.183i\\ 0 - i \end{array}\right)&
\left (\begin{array}{c}-0.683 +	0.183i\\0.183 - 0.683i\\ 0 \end{array}\right)&
\left (\begin{array}{c}0.183 +	0.683i\\0.683 +	0.183i\\ 0 \end{array}\right)\\
&&&&\\
\psi^{1}_{12}(1,0,0) &
\left (\begin{array}{c} 0 \\-0.183 - 0.683i\\ -0.683 - 0.183i \end{array}\right)&
\left (\begin{array}{c} 0 \\-0.683 + 0.183i\\ -0.183 + 0.683i \end{array}\right)&
\left (\begin{array}{c} 1\\-0.683 + 0.183i\\ 0.683 + 0.183i \end{array}\right)&
\left (\begin{array}{c} 0 - i\\-0.183 - 0.683i\\ -0.183 + 0.683i\\ \end{array}\right)\\
&&&&\\
\psi^{1}_{12}(0,1,0)  &
\left (\begin{array}{c} -0.683 - 0.183i\\ 0 \\ -0.183 - 0.683i\\ \end{array}\right)&
\left (\begin{array}{c} 0.183 - 0.683i\\ 0 \\ -0.683 + 0.183i\\ \end{array}\right)&
\left (\begin{array}{c} 0.683 +	0.183i\\ -1 \\ 0.683 - 0.183i\\ \end{array}\right)&
\left (\begin{array}{c} 0.183 - 0.683i\\0 - i \\ 0.183 + 0.683i\\ \end{array}\right)\\
&&&&\\
\psi^{1}_{12}(0,0,1) &
\left (\begin{array}{c}-0.183	- 0.683i\\-0.683 - 0.183i\\ 0  \end{array}\right)&
\left (\begin{array}{c} 0.683 - 0.183i\\-0.183 +	0.683i\\ 0 \end{array}\right)&
\left (\begin{array}{c} 0.683 - 0.183i\\-0.683 - 0.183i\\ 1 \end{array}\right)&
\left (\begin{array}{c}-0.183	- 0.683i\\0.183	- 0.683i\\ 0 + i \end{array}\right)\\
\end{array}$}
\hfill{}
\label{tb:BVTb1}
\end{table*}
\end{center}

\begin{center}
\begin{table*}[ht]
\caption{The BVs projected for our example system using symmetry adapted trial functions.}
{\small
\hfill{}
$\begin{array}{l|cccc}      
BV & A_{0} = (0,0,0) & A_{1} = (\frac{1}{2}, \frac{1}{2}, 0) &A_{2} = (0, \frac{1}{2}, \frac{1}{2})& A_{3} = (\frac{1}{2}, 0, \frac{1}{2}) \\
\hline
&&&&\\
\frac{1}{\sqrt{3}}\psi^{1}_{11}(1,1,1)&
\left (\begin{array}{c}.366\\.366\\.366 \end{array}\right)&
\left (\begin{array}{c}.366i\\-.366i\\ -.366i \end{array}\right)&
\left (\begin{array}{c} - 0.5 + 0.5i\\0.5 - 0.5i\\-0.5 + 0.5i \end{array}\right)&
\left (\begin{array}{c} 0.5 + 0.5i\\0.5 + 0.5i\\ -0.5 - 0.5i \end{array}\right)\\
&&&&\\
\frac{1}{\sqrt{2}}\psi^{1}_{11}(1,-1,0)&
\left (\begin{array}{c}0.837 - 0.483i\\-0.837 - 0.483i\\ 0 + 0.966i \end{array}\right)&
\left (\begin{array}{c}0.483 + 0.837i\\-0.483 + 0.837i\\ 0.966 \end{array}\right)&
\left (\begin{array}{c} 0.129 - 0.483i \\ 0.483 - 0.129i\\ 0.354 + 0.354i \end{array}\right)&
\left (\begin{array}{c} -0.483 - 0.129i \\0.129 + 0.483i\\ -0.354 + 0.354i \end{array}\right)\\
&&&&\\
\frac{1}{\sqrt{6}}\psi^{1}_{11}(1,1,-2)&
\left (\begin{array}{c} 0.483 + 0.837i\\0.483 - 0.837i\\ -0.966 \end{array}\right)&
\left (\begin{array}{c} -0.837 + 0.483i\\-0.837 - 0.483i\\ 0 + 0.966i \end{array}\right)&
\left (\begin{array}{c} 0.483 + 0.129i \\0.129 + 0.483i\\ -0.354 + 0.354i \end{array}\right)&
\left (\begin{array}{c} 0.129 - 0.483i \\ -0.483 + 0.129i\\ -0.354 - 0.354i \end{array}\right)\\
\end{array}$}
\hfill{}
\label{tb:BVTb2}
\end{table*}
\end{center}

\begin{center}
\begin{table*}[ht]
\caption{The BVs projected for the second orbit of our example system using symmetry adapted trial functions.}
{\small
\hfill{}
$\begin{array}{l|cccc}      
BV & A_{4} = (\frac{1}{4},\frac{3}{4},\frac{3}{4}) & A_{5} = (\frac{3}{4},\frac{3}{4},\frac{1}{4}) &A_{6} = (\frac{3}{4},\frac{1}{4},\frac{3}{4})& A_{7} = (\frac{1}{4},\frac{1}{4},\frac{1}{4}) \\
\hline
&&&&\\
\frac{1}{\sqrt{3}}\psi^{1}_{11}(1,1,-1)&
\left (\begin{array}{c} 1.366\\1.366\\-1.366 \end{array}\right)&
\left (\begin{array}{c} 1.366i\\-1.366i\\ 1.366i \end{array}\right)&
\left (\begin{array}{c}-0.5 - 0.5i\\ 0.5 + 0.5i\\ 0.5 + 0.5i \end{array}\right)&
\left (\begin{array}{c}-0.5 + 0.5i\\-0.5 + 0.5i\\-0.5 + 0.5i \end{array}\right)\\
&&&&\\
\frac{1}{\sqrt{2}}\psi^{1}_{11}(-1,1,0)&
\left (\begin{array}{c} -0.224 + 0.129i \\ 0.224 + 0.129i\\ 0.259i \end{array}\right)&
\left (\begin{array}{c} -0.129 - 0.224i \\ 0.129 - 0.224i\\ 0.259 \end{array}\right)&
\left (\begin{array}{c} -0.483 - 0.129i \\-0.129 - 0.483i\\-0.354 + 0.354i \end{array}\right)&
\left (\begin{array}{c} -0.129 + 0.483i \\ 0.483 - 0.129i\\-0.354 - 0.354i \end{array}\right)\\
&&&&\\
\frac{1}{\sqrt{6}}\psi^{1}_{11}(1,1,2)&
\left (\begin{array}{c} 0.129 + 0.224i \\ 0.129 - 0.224i\\ 0.259 \end{array}\right)&
\left (\begin{array}{c}-0.224 + 0.129i \\-0.224 + 0.129i\\-0.259i \end{array}\right)&
\left (\begin{array}{c}-0.129 + 0.483i \\-0.483 + 0.129i\\+0.354 + 0.354i \end{array}\right)&
\left (\begin{array}{c} 0.483 + 0.129i \\-0.129 + 0.483i\\-0.354 + 0.354i \end{array}\right) \\
\end{array}$}
\hfill{}
\label{tb:BVTb3}
\end{table*}
\end{center}

\end{document}